\documentclass[%
 reprint,
superscriptaddress,
showpacs,
 amsmath,amssymb,
 aps,prd
]{revtex4-1}

\usepackage{graphicx}
\usepackage{dcolumn}
\usepackage{bm}
\usepackage[mathlines]{lineno}
\usepackage{enumitem}
\usepackage{rotating}
\usepackage{epsfig}
\usepackage{natbib}
\usepackage{color}
\usepackage{datetime}
\usepackage{wrapfig}
\usepackage{threeparttable}
\usepackage{color}

\def\bi{\begin{itemize}[noitemsep,leftmargin=*]}
\def\ei{\end{itemize}}

\usepackage{soul}

\newcommand\chandra{{\it Chandra}}
\newcommand\xmm{{\it XMM-Newton}}
\newcommand\suzaku{{\it Suzaku}}
\newcommand\integral{{\it INTEGRAL}}

\newcommand\nustar{\hbox{\it NuSTAR}}

\newcommand\cobe{{\it COBE/DIRBE}}

\newcommand{\smallket}[1]{ \left(#1 \right)}

\newcommand\T{\rule{0pt}{2.6ex}}       
\newcommand\B{\rule[-1.2ex]{0pt}{0pt}}

\def\arcmin{\hbox{$^\prime$}}
\def\arcsec{\hbox{$^{\prime\prime}$}}

\begin{document}

\preprint{APS/123-QED}

\title{(Almost) Closing the $\nu$MSM Sterile Neutrino Dark Matter Window with \emph{NuSTAR} }

\author{Kerstin Perez}
\email{kmperez@mit.edu}
\affiliation{Department of Physics, Massachusetts Institute of Technology, Cambridge, MA 02139, USA}
\affiliation{Department of Physics, Haverford College, Haverford, PA 19141, USA}%

\author{Kenny C. Y. Ng}
\thanks{ng.199@osu.edu}
\affiliation{Center for Cosmology and AstroParticle Physics (CCAPP), Ohio State University, Columbus, OH 43210, USA}
\affiliation{Department of Physics, Ohio State University, Columbus, OH 43210, USA}

\author{John F. Beacom}
\email{beacom.7@osu.edu}
\affiliation{Center for Cosmology and AstroParticle Physics (CCAPP), Ohio State University, Columbus, OH 43210, USA}
\affiliation{Department of Physics, Ohio State University, Columbus, OH 43210, USA}
\affiliation{Department of Astronomy, Ohio State University, Columbus, OH 43210, USA} 

\author{Cora Hersh}
\email{chersh@haverford.edu }
\affiliation{Department of Physics, Haverford College, Haverford, PA 19141, USA}

\author{Shunsaku Horiuchi}
\email{horiuchi@vt.edu}
\affiliation{Center for Neutrino Physics, Department of Physics, Virginia Tech, Blacksburg, VA 24061, USA}%

\author{Roman Krivonos}
\email{krivonos@iki.rssi.ru}
\affiliation{Space Research Institute of the Russian Academy of Sciences (IKI) Moscow, Russia, 117997}%

\date{20th November, 2016}

\begin{abstract}

We use \nustar\ observations of the Galactic Center to search for X-ray lines from the radiative decay of sterile neutrino dark matter. Finding no evidence of unknown lines, we set limits on the sterile neutrino mass and mixing angle. In most of the mass range 10--50\,keV, these are now the strongest limits, at some masses improving upon previous limits by a factor of $\sim10$.  
In the $\nu$MSM framework, where additional constraints from dark matter production and structure formation apply, the allowed parameter space is reduced by more than half. Future \nustar\ observations may be able to cover much of the remaining parameter space. 
\end{abstract}


\maketitle

\begin{figure}[t!]
\vspace{0.1cm}
\includegraphics[width=1.0\linewidth]{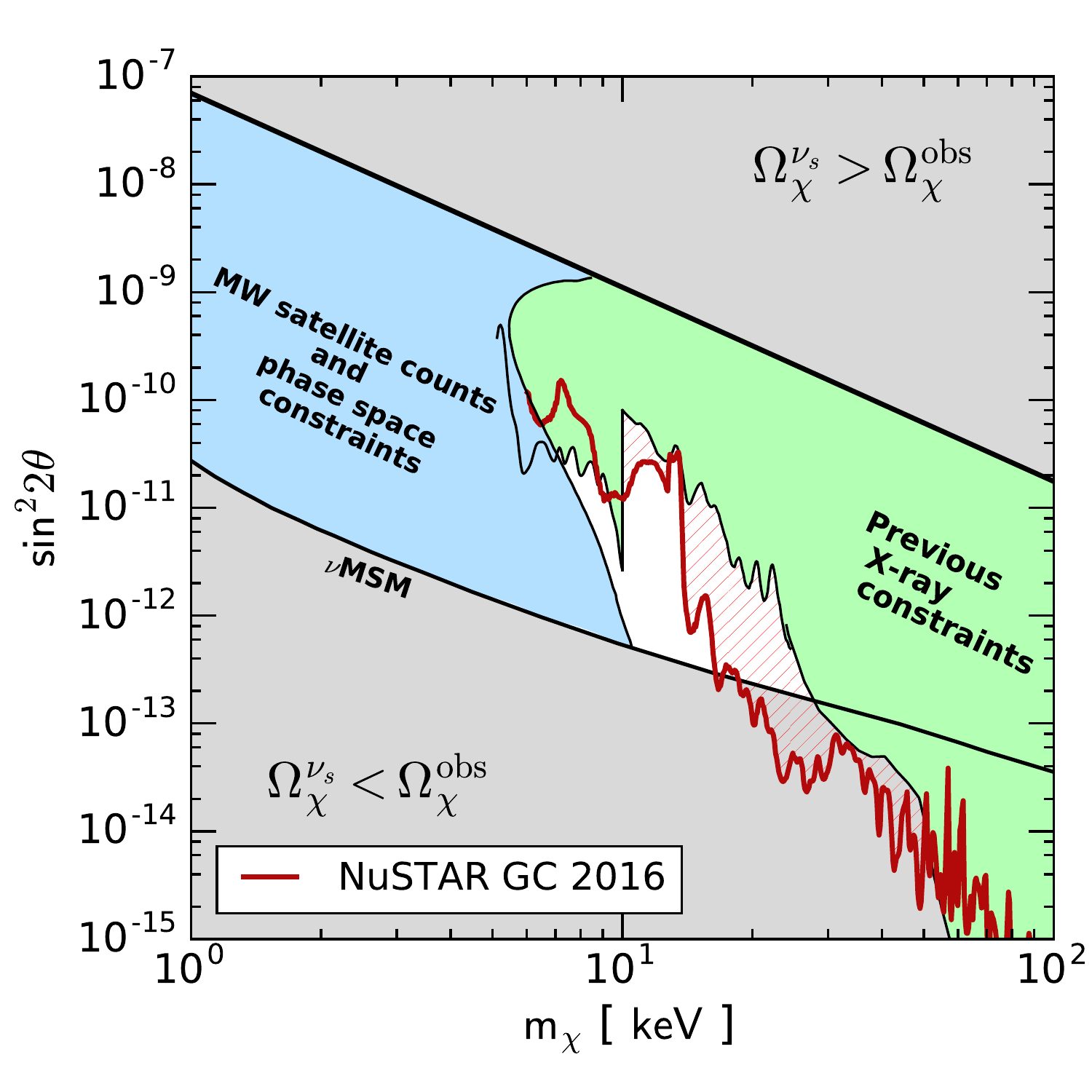}
\caption{\label{fig:intro}
Simplified overview of constraints on $\nu$MSM sterile neutrino dark matter in the plane of mass and mixing angle; details are described in Sec.~\ref{sec:results}. For parameters between the solid black lines, the observed dark matter abundance can be produced through resonant production in the $\nu$MSM. 
Most of this region has been ruled out by constraints from structure formation considerations (blue region) or astrophysical X-ray observations (green region).  Our new constraint (red line and hatched region) is obtained from \nustar\ observations of the GC, and rules out about half of the previously allowed parameter space (white region).}
\end{figure}

\section{\label{sec:intro} Introduction}

Is dark matter composed entirely of sterile neutrinos?  In the neutrino minimal standard model ($\nu$MSM~\cite{Asaka:2005an, Boyarsky:2006jm, Asaka:2006nq, Canetti:2012vf, Canetti:2012kh}) --- an economical framework that can simultaneously explain neutrino mass, the baryon asymmetry of the Universe, and dark matter --- a definitive answer is possible.  
Sterile neutrino dark matter can be produced through mixing with active neutrinos.  
In the $\nu$MSM, where the lepton asymmetry is non-zero, dark matter is produced with resonant production, also called the Shi-Fuller mechanism~\cite{Shi:1998km}.  (In the limit of zero lepton asymmetry, it corresponds to non-resonant production, also called the Dodelson-Widrow mechanism~\cite{Dodelson:1993je}.)
When all constraints are considered, the allowed parameter space for sterile neutrino dark matter in the $\nu$MSM is finite~(see Refs.~\cite{Boyarsky:2009ix, Boyarsky:2012rt, Adhikari:2016bei} for reviews).

In Fig.~\ref{fig:intro}, we summarize the current constraints and the improvements resulting from the work presented in this paper~(detailed in Sec.~\ref{sec:results}). Astrophysical X-ray constraints are model independent and provide \emph{upper} limits on the sterile neutrino mass~\cite{Dolgov:2000ew, Abazajian:2001vt}.  If the $\nu$MSM is considered, structure-formation considerations provide \emph{ lower} limits on the mass~\cite{Dodelson:1993je, Shi:1998km,Dolgov:2000ew, Abazajian:2001nj}.
At smaller masses~($\lesssim 10$\,keV), there are strong limits from X-ray telescopes such as \chandra, \suzaku, and \xmm, while at larger masses~($\gtrsim 50$\,keV), there are strong limits from \integral.
However, until now, it has been particularly difficult to probe masses in the range 10--50 keV, which, since radiative decay produces an X-ray line at energy $E_\gamma = m_{\chi}/2$, corresponds to X-rays of energies 5--25 keV.  This has been mostly due to the lack of new instruments sensitive to the relevant X-ray energy range.

\begin{table*}
\caption{\label{tab:obs} \nustar\ observations used for this analysis.}
\begin{ruledtabular}
\begin{tabular}{ccccccc}
\multicolumn{1}{c}{Observation ID} & \multicolumn{2}{c}{Pointing (J2000)\footnote{Roll angle was 332$^\circ$ for all.} } & Effective Exposure\footnote{After all data cleaning.} & Detector Area\footnote{After stray light, ghost ray, and bad pixel removal.} & Avg. Solid Angle\footnote{Average solid angle of sky from which 0-bounce photons can be detected, after correcting for removal of stray light, ghost rays, and bad pixels, as well as efficiency due to vignetting effects.} \\
\multicolumn{1}{r}{} 		& RA (deg) & DEC (deg) & FPMA\,/\,FPMB (ks)  & FPMA\,/\,FPMB (cm$^2$)       & FPMA\,/\,FPMB (deg$^2$) \B \\
\hline
\multicolumn{1}{r}{40032001002}  	& 265.8947 & $-29.5664$ & 39.7\,/\,39.6 & 9.89\,/\,11.10 & 3.73\,/\,4.09 \T \\
\multicolumn{1}{r}{40032002001}  				& 265.7969 & $-29.5139$ & 39.8\,/\,39.6 & 7.14\,/\,8.05 & 4.06\,/\,4.12 \\
\multicolumn{1}{r}{40032003001} 					& 265.6991 & $-29.4613$ & 39.8\,/\,39.6 & 8.18\,/\,8.92 & 3.47\,/\,4.01 \\
\multicolumn{1}{r}{40032004002}  				& 265.9550 & $-29.4812$ & 22.6\,/\,22.7 & 4.19\,/\,6.54 & 2.34\,/\,3.13 \\
\multicolumn{1}{r}{40032005002} 					& 265.8572 & $-29.4288$ & 25.6\,/\,25.8  & 9.78\,/\,7.85 & 3.80\,/\,3.85 \\
\multicolumn{1}{r}{40032006001} 					& 265.7595 & $-29.3762$ & 28.6\,/\,28.6 & 9.98\,/\,6.18 & 3.76\,/\,3.74 \\
\end{tabular}
\end{ruledtabular}
\end{table*}

Launched in 2012, the \emph{Nuclear Spectroscopic Telescope Array} (\nustar)~\cite{Harrison:2013md} is the first focusing optic to cover the 3--79~keV energy range. 
Due to its combination of grazing-incidence design and multilayer-coated reflective optics, \nustar\ provides unprecedented sensitivity in this hard X-ray band, and its focal-plane detectors deliver energy resolution of 400~eV at $E_\gamma = 10$~keV. Moreover, \nustar\ has already completed (\emph{i.}) long exposures of the Galactic Center (GC), where the dark matter decay signal is expected to be bright, as well as (\emph{ii.}) extensive modeling of the astrophysical emission components, which form a significant background to sterile neutrino searches~\cite{NuSTARGRXE}.

Due to the geometry of the \nustar\ instrument, photons arriving from several degrees away from the target of observation may directly enter the detectors without passing through the focusing optics.  These ``0-bounce" photons (see Sec.~\ref{sec:obs}) normally constitute a background for pointed observations.  However, an innovative use of these photons is to probe large-scale diffuse emission that extends over much larger scales than the field of view (FOV) of focused photons.  We exploit the wide \nustar\ solid angle aperture for 0-bounce photons to perform a sensitive search for dark matter decay in the GC region.  As show in Fig.~\ref{fig:intro}, this reduces the remaining parameter space for sterile neutrino dark matter in the $\nu$MSM by about half.  

In Sec.~\ref{sec:nustaranalysis}, we describe the \nustar\ instrument and the dataset used in this analysis (Sec.~\ref{sec:obs}), the particular analysis procedures necessary to utilize 0-bounce photons (Sec.~\ref{sec:zerobounce}), and the energy spectrum of the GC and corresponding line-search analysis (Sec.~\ref{sec:spec}). In Sec.~\ref{sec:jfactor}, we model the expected dark matter signal, which takes into account the non-trivial shape of the aperture for 0-bounce photons.  In Sec.~\ref{sec:results}, we present our results in the mass-mixing plane and put them in the context of previous constraints.  Conclusions and comments on future prospects are presented in Sec.~\ref{sec:conclusion}.

\begin{figure}
\includegraphics[width=0.75\linewidth]{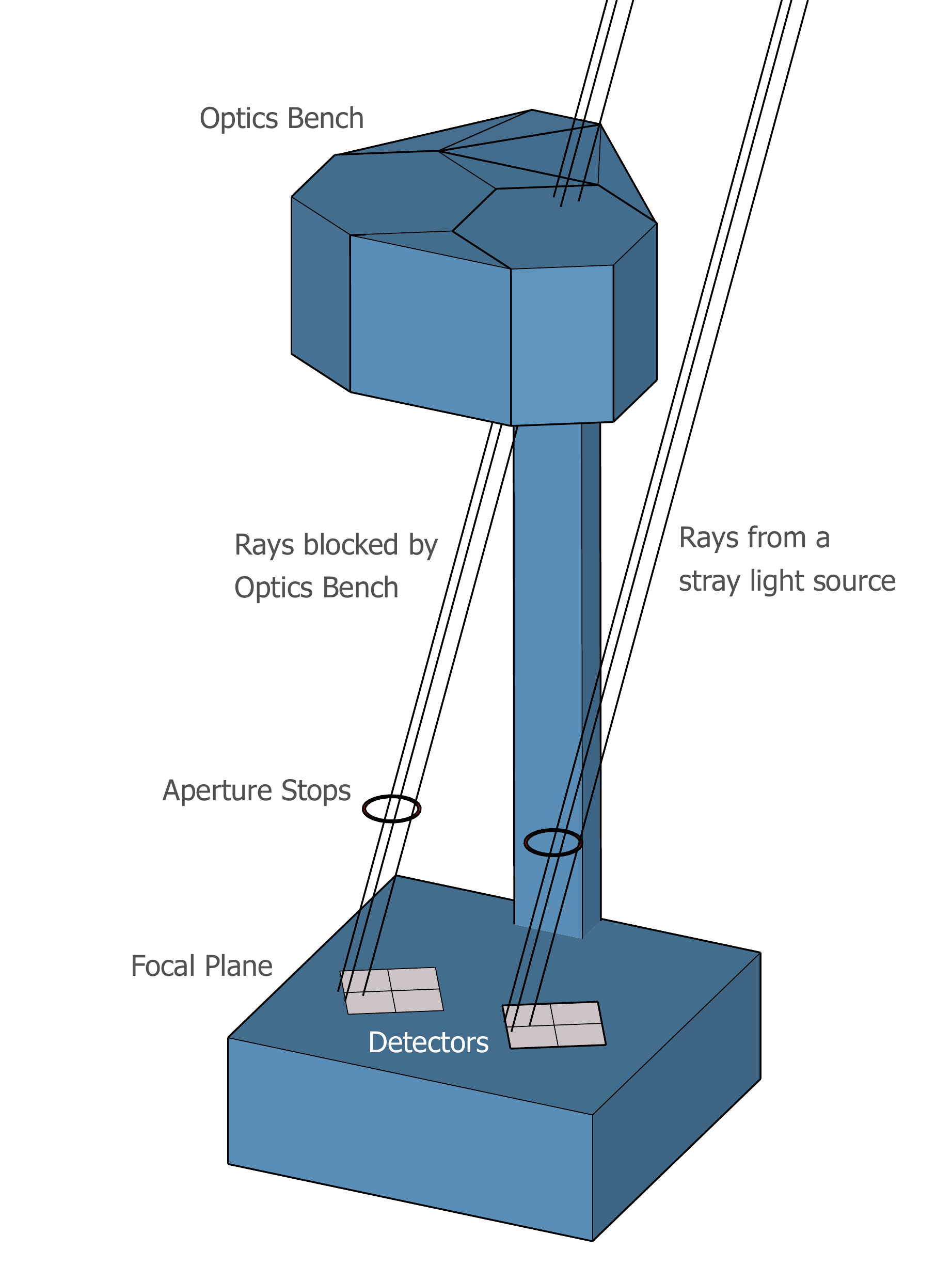}
\caption{\label{fig:0bounce} Illustration, from Ref.~\cite{Wik:2014boa}, of the \nustar\ observatory geometry. 0-bounce photons from far off-axis sources can bypass the aperture stops and shine directly on the detectors, though some of these rays are blocked by the optics bench.}
\end{figure}

\section{\label{sec:nustaranalysis} NuSTAR Data Analysis}

\subsection{\label{sec:obs} NuSTAR instrument and GC observations}
 
\emph{NuSTAR} has two identical telescopes, each consisting of an independent optic and focal-plane detector, referred to as FPMA and FPMB. The optics use a multilayer coating combined with a conical approximation to the grazing-incidence Wolter-I design, in which X-rays reflect from an upper parabolic mirror section and then a lower hyperbolic mirror section, to focus photons in the energy range 3--79\,keV. Each focal-plane module has a FOV for focused (``2-bounce") X-rays of $13\arcmin\ \times 13\arcmin\ $. 

To block unfocused X-rays from reaching the \nustar\ detectors, a series of aperture stops are attached to each focal-plane bench. 
Still, this shielding is not complete, and there remains a $\sim5^\circ$-radius aperture, partially blocked by the optics bench, from which totally unfocused, or ``0-bounce", photons can reach the detectors~(see Fig.~\ref{fig:0bounce}). 
In addition, photons arriving from within $\sim1^\circ$ of the optical axis can reflect once from only the upper or lower mirror section, and are known as ``1-bounce" photons or ghost rays. 

\emph{NuSTAR} performed pointed observations of the central $\sim1.4^\circ\times0.6^\circ$ of the Galaxy for a total of $\sim2$~Ms over the period from July 2012 through October 2014~\cite{Mori:2015vba,Hong:2016}. 
We use six tiled observations (Table~\ref{tab:obs}), chosen to minimize flux from bright sources closer to the GC, from the ``Block B" survey~\cite{Hong:2016}. 
Data reduction and spectral extraction were performed with the \nustar\ Data Analysis Software pipeline (\emph{NuSTARDAS}) v1.5.1. 

We remove all data taken during passage through the South Atlantic Anomaly (SAA). 
Using a geometric model of the telescope, we flag as ``bad" any pixels that have significant contamination from 0-bounce photons caused by bright, localized sources at large off-axis angles (known as stray light~\cite{Harrison:2013md, Krivonos:2013sqa, Wik:2014boa}).   
These pixels are then removed during the data screening procedure implemented in \emph{NuSTARDAS}.
Ghost rays that are caused by these bright sources can produce high-intensity radial streaks in the image~\cite{Mori:2015vba}, and are also removed during data screening. 
We do not remove 2-bounce photons from known point sources (except for a 15\arcsec\, radius around the bright source 174306.9-292327~\cite{Muno2009,Hong:2016}), as their contamination is negligible (see Sec.~\ref{sec:zerobounce} and \ref{sec:spec}). 
Spectra are extracted from all remaining detector regions. 

\begin{figure}
\includegraphics[width=1.0\linewidth]{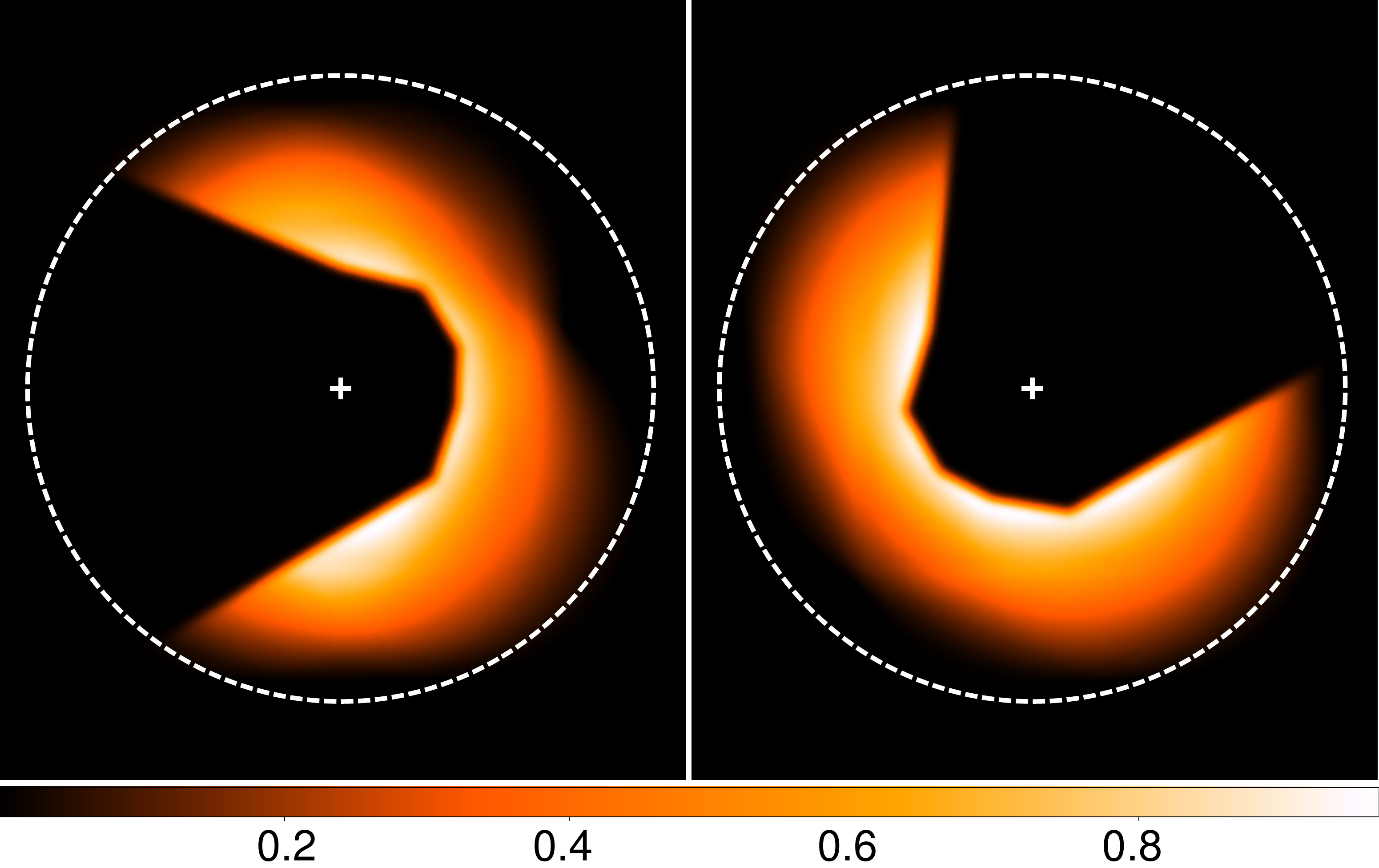}
\caption{\label{fig:pacman} Aperture area for 0-bounce photons detected by FPMA (left) and FPMB (right) in Observation ID 40032001002. The detector efficiency due to vignetting effects is indicated by the color scale (arbitrary units). The white dashed line shows a 3.5$^\circ$-radius around the pointing of this observation, indicated by the cross. The optical bench structure obscures the triangular region, resulting in the ``Pac-Man" shape. }
\end{figure}

\subsection{\label{sec:zerobounce} Spectral analysis with 0-bounce photons}

\begin{figure*}
\includegraphics[width=0.9\linewidth]{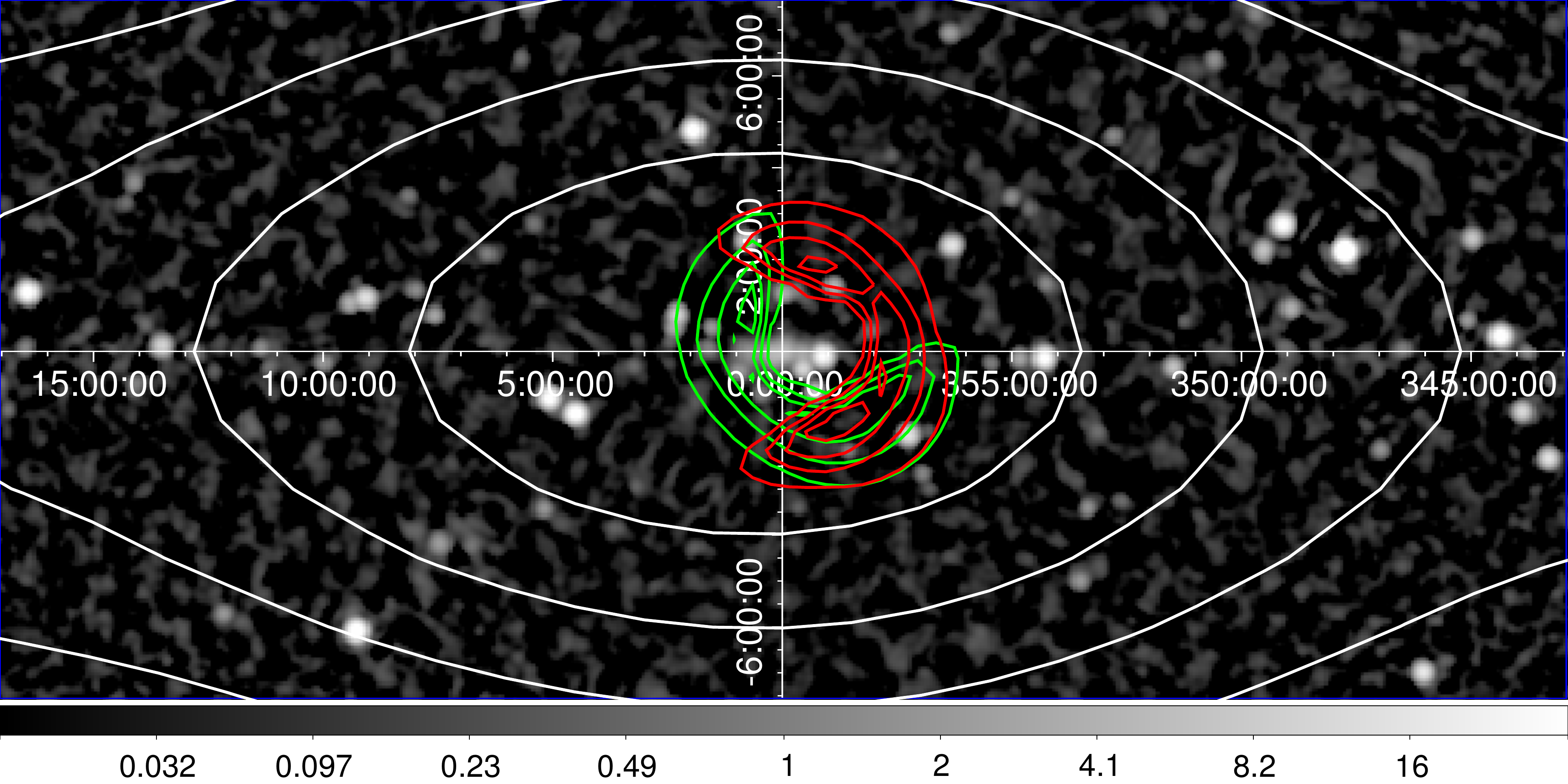}
\vspace{-0.in}
\caption{\label{fig:sky} Sky coverage of 0-bounce photons from FPMA (red) and FPMB (green) after removal of stray light, bad pixels, and ghost rays, as well as correcting for efficiency due to vignetting effects, overlaid on the 9-year \integral\ image of the central $30^\circ \times 12^\circ$ of the Galaxy in 17--60\,keV~\cite{2012A&A...545A..27K}. The gray color scale is in units of mCrab. The red and green contours indicate the efficiency due to vignetting effects. The projected stellar mass density distribution, as traced by the 4.9-$\mu$m surface brightness measured by \cobe\ (provided by the LAMBDA archive of the Goddard Space Flight Center, http://lambda.gsfc.nasa.gov), is indicated by the white contours. }
\end{figure*}

Because the solid-angle aperture for 0-bounce photons is over two orders of magnitude larger than the FOV for focused photons, observations of diffuse emission that extends over many degrees will be dominated by the 0-bounce flux. 
The six observations we use thus have a count rate dominated by 0-bounce photons, because the underlying Galactic Ridge X-ray Emission~(GRXE)~\cite{Worrall1982,Kaneda1997,ValiniaMarshall1998,Revnivtsev2006,Krivonos2007,Yuasa2012} extends for tens of degrees along the Galactic plane. 
Even after accounting for the increase in X-ray intensity toward the more central region covered by the 2-bounce FOV, our spectrum still contains more than an order of magnitude more 0-bounce photons than 2-bounce photons.
We note that this is not a problem for analyses of point sources in this region, because the 0-bounce contribution can be subtracted using spectra from nearby ``empty" (i.e., diffuse dominated) regions. 

The use of 0-bounce photons for our spectral analysis has several implications. 
The main disadvantage is a lower effective area for the telescopes. 
The effective area for focused photons is determined mainly by the \nustar\ optics, each of which have an effective area of $\sim$\,1000\,cm$^2$ at 10\,keV and $\sim$\,200\,cm$^2$ at 40\,keV~\cite{Harrison:2013md}; the effective area for 0-bounce photons is determined mainly by the physical detector area, which is only $\sim$\,15\,cm$^2$ per module. 
This is balanced, however, by two large advantages.
First, since 0-bounce photons arrive from a much larger sky area, we expect a larger flux from dark matter decays~(see Sec.~\ref{sec:jfactor}). 
Second, we are not constrained to the energy range of the optics, so we can use the larger energy range of the focal-plane detectors, $E_\gamma = $ 3--110\,keV. 

To search for sterile neutrino dark matter, we need the true sky area (in units of deg$^2$) that the 0-bounce photons in our spectrum are coming from. 
We use the \emph{nuskybgd} code~\cite{Wik:2014boa} to construct a sky-exposure map for each observation, corrected for the vignetting effect produced by the aperture stop and obscuration by the optical bench structure, producing the ``Pac-Man" shape shown in Fig.~\ref{fig:pacman}. After all data cleaning, this solid-angle aperture for 0-bounce photons has a radius of $\sim3.5^\circ$.
The combined sky coverage of 0-bounce photons from FPMA and FPMB for all six of our observations is shown in Fig.~\ref{fig:sky}.  

We normalize each individual observation spectrum to (\emph{i.}) the physical detector area that remains after removing bad pixels, stray light, and ghost rays, and (\emph{ii.}) the 0-bounce aperture area in units of deg$^{2}$, using the values listed in Table~\ref{tab:obs}. 
The spectra of the six observations are then combined separately for FPMA and FPMB, and normalized to the exposure-time weighted average effective detector area and exposure-time weighted average solid angle of sky coverage. 
This yields a spectrum in units of $\rm ph\,cm^{-2}\,s^{-1}\,deg^{-2}\,keV^{-1}$.

\subsection{\label{sec:spec} Spectral fit and line analysis}

Any search for a line feature in an astrophysical spectrum will be limited by the statistical and systematic uncertainty of the measured spectrum, as well as the energy resolution of the instrument.

By using 0-bounce photons, we have over $10^5$ photons from each of FPMA and FPMB, in the energy range 3--110\,keV. 
With this large number of total counts in each spectrum, we have the flexibility to choose a binning scheme that is optimized to be both narrow enough to distinguish spectral features, but also wide enough to minimize the statistical uncertainty of each bin. 
The spectra for FPMA and FPMB are each binned using a logarithmic binning scheme with 200 bins per decade.
This is chosen so that each bin in the energy range of interest is narrower than the one-photon \nustar\ energy resolution (FWHM), which varies from 400\,eV at 10\,keV to 900\,eV at 60\,keV~\cite{Harrison:2013md}, and also wider than the many-photon energy resolution, $\sim$FWHM/$\sqrt N$. 
With this choice of binning, the spectrum from each module has $\sim$\,600 photons per bin at the lowest energies and $\sim$\,350 photons per bin at the highest energies, providing a statistical uncertainty that is everywhere $\sim$\,4--5\%. 
This binning scheme also allows for easy visual display of relevant spectral features. 
We observe no significant variations in the underlying model flux or the derived maximum sterile neutrino flux for alternative binning schemes with linear widths ranging from 40--160\,eV. 

\begin{figure*}
\includegraphics[width = 2\columnwidth]{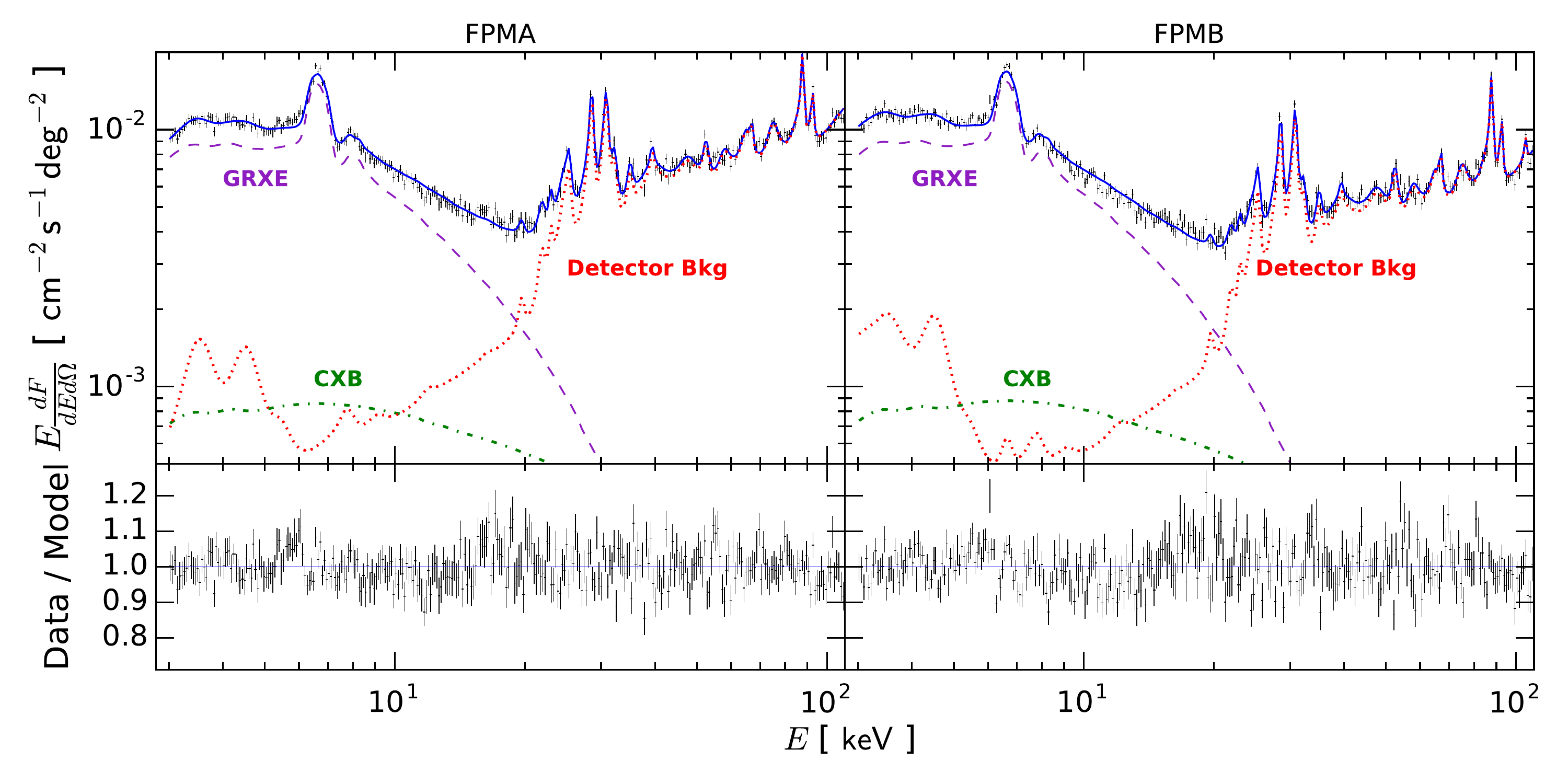}
\vspace{-0.2in}
\caption{\label{fig:spectrum} Data and folded model spectra from FPMA (left) and FPMB (right) in 3--110\,keV. Model components include the GXRE~(line and continuum), the CXB (continuum), and detector backgrounds~(line and continuum).  The astrophysical components come from regions indicated in Fig.~\ref{fig:sky}. The bottom panel shows the data relative to the best-fit model. All errors shown are 1$\sigma$ statistical errors. We include an additional 5\% uncorrelated systematic error~(not shown) during spectral fitting and line analysis. 
}
\end{figure*}

There are also systematic uncertainties, such as that arising from the use of one set of model parameters to describe an astrophysical background that varies slightly between each observation region.
These differences in the underlying source population can cause a change not only in the overall flux value of the astrophysical background, but also in the shape of this background spectrum.
Other sources of systematic uncertainty could come from non-uniform detector response as a function of energy.
We assign a 5\% systematic error, conservatively taken to be uncorrelated bin-to-bin, in order to account for these effects. 
This choice of systematic error minimizes the fit residuals and yields a $\chi^2$/n.d.o.f $\approx 1$.

In addition to this uncorrelated systematic error, which is included during fit optimization, there is an overall \nustar\ flux normalization uncertainty, which is not included. 
By comparing to other X-ray instruments, the overall flux normalization uncertainty has been experimentally determined to be $\sim$\,10\%~\cite{Madsen:2015jea}. 
This additional uncertainty only shifts the overall flux limit by $\sim$\,10\%, which is negligible compared to other sources of uncertainties, such as the Milky Way dark matter content.

We do not co-add the two spectra from FPMA and FPMB, due to differences in the internal detector background spectrum and in the overall flux normalizations for each focal-plane module.
Instead, we perform simultaneous fitting of the two spectra, where all astrophysical parameters are constrained to be the same for each focal-plane module, but all internal detector background parameters are fit individually. 
A floating constant factor is included in our spectral model to account for the different flux normalizations. For our best-fit model, this factor is $<3$\%, smaller than the overall \nustar\ flux normalization uncertainty. 
The fluxes we quote below are derived for FPMA.

Our spectral model consists of four components, two from astrophysical sources and two internal to the detector. 
The GRXE, believed to be largely due to unresolved magnetic cataclysmic variables~\cite{Revnivtsev2006,Krivonos2007,Yuasa2012}, is modeled as a one-temperature thermal plasma with collisionally-ionized elemental line emission~\cite{Smith2001}, which describes the X-ray emitting accretion stream onto these objects, plus a 6.4~keV neutral Fe line, with the normalization of the Gaussian line and the normalization, temperature, and abundance of the plasma left as free parameters.
Using the NuSTAR GC source catalog~\cite{Hong:2016}, the total 10--40 keV flux of resolved 2-bounce sources in our FOV is $\sim10^{-6}$~ph~s$^{-1}$~cm$^{-2}$. This negligibly small contribution of flux is absorbed into our GRXE model.
The temperature of the GRXE in this one-temperature model varies by up to 20\% between the six observations, motivating the uncorrelated systematic error that is included in our fit of the combined spectrum.
The cosmic X-ray background (CXB), due to extragalactic emission, is modeled as a cutoff power-law, with parameters fixed to those measured by \integral~\cite{Churazov:2006bk}. 
These spectra are attenuated to account for absorption by the interstellar medium, with interstellar abundances as defined in~\cite{1989GeCoA..53..197A} and photoionization cross-sections as defined in~\cite{1992ApJ...400..699B,0004-637X-496-2-1044}. 
The effective area for these two model components, which describe photons arriving from astrophysical sources, is multiplied by the energy-dependent efficiency for photons to pass through the detector beryllium shield.
All model components include an absorption term that accounts for detector focal-plane material.

\begin{figure*}[]
\begin{centering}
\includegraphics[width=2.0\columnwidth]{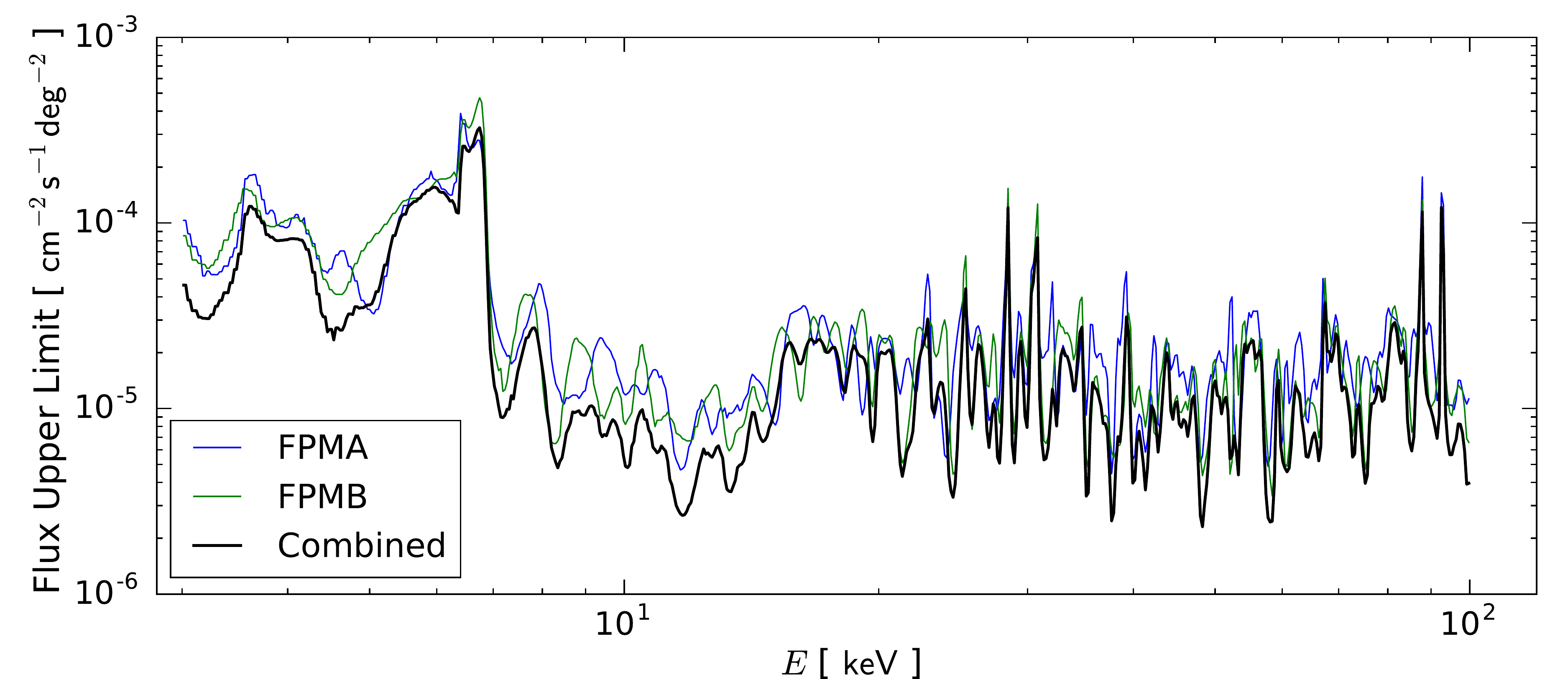}
\caption{\label{fig:maxflux} Flux upper limits (one-sided 95\% C.L.) for the normalizations of possible line signals.  These are derived from Fig.~\ref{fig:spectrum}  taking into account the allowed excesses over background in appropriately narrow ranges of bins.  We show limits for the FPMA (blue) and FPMB (green) detectors, which are independent, as well as for their combination (black). }
\end{centering}
\end{figure*}

The internal detector background consists of a continuum component, modeled as a broken power-law with a break at 124~keV, and both activation and fluorescent line complexes, modeled as 29 Lorentzian lines~\cite{Wik:2014boa}. 
The continuum photon indices and line energies are fixed, but normalizations for each component are fit separately for FPMA and FPMB.  
Since these components describe backgrounds that are internal to the detectors, they are not corrected for the efficiency of the beryllium shield.
The solar background, modeled as a $\sim1$~keV thermal plasma as derived in~\cite{Wik:2014boa}, is also included in this component.

In Fig.~\ref{fig:spectrum} we show the 3--110~keV data and folded best-fit spectral model for FPMA and FPMB, respectively.  
This model contains 69 free parameters and 45 frozen parameters, with the fit performed over $312 \times 2$~(FPMA and FPMB) total bins.
We emphasize that these two data sets are independent of each other; our results are obtained by statistically combining them.
Spectral fitting and flux derivations were performed in XSPEC version 12.9.0~\cite{Arnaud1996}.
The combined fit yields a $\chi^2 = 540.02$ for 554 degrees of freedom, or $\chi^2$/n.d.o.f.$ = 0.97$ (both statistical and 5\% systematic errors included). 
The physical interpretation of the best-fit GRXE spectrum will be the subject of a future paper, and is not important for this analysis.
The critical quantity for the current analysis is the quality of the fit to the spectrum. 

We search for a possible component of emission due to sterile neutrinos by adding to the above model a line at a fixed energy. 
For this added line signal, we take into account all detector effects, such as energy resolution and absorption from the detector beryllium shield, as well as astrophysical effects, such as absorption from the interstellar medium.
We scan for signals with line energies of 3--110~keV in logarithmic energy steps of 400 steps per decade. 
No new line excess is found in the search.

We then proceed to set exclusion limits based on the null result.
For each fixed dark matter line energy, we first vary all the model parameters~(including the dark matter normalization) and find the best-fit values by minimizing the $\chi^2$.  We then increase the dark matter line normalization starting from the best-fit value, while at each step allowing for simultaneous variations of all other free model parameters, until the $\chi^2$ varies from its best-fit value by $\Delta\chi^2 = 2.71$.  During the procedure, we allow all line normalizations to be non-zero.
This is a conservative method to derive a limit, as it allows the dark matter line to assume the full strength of any known astrophysical or detector background line.
This corresponds to a one-sided 95\% C.L. upper limit on the sterile neutrino line flux~\cite{Rolke:2004mj, Cowan:2010js}.

In Fig.~\ref{fig:maxflux} we show the derived maximum line flux as a function of photon energy.  The expected limit can be roughly estimated by considering the two major factors affecting the line analysis in an energy range where the spectrum is well fit by a continuum model, e.g., around 10\,keV.  The first is the line width, set by the detector energy resolution, which has a value of roughly $\delta E/E \sim \delta \log E \sim 5\%$.  The second is the maximum allowed contribution of the line to the model, set roughly by the total error of each data bin, which has a value of $\sim 5\%$.  
These two factors combined explain the relative factor of $\sim 10^{-3}$ between the total flux shown in Fig.~\ref{fig:spectrum} and the derived line limit shown in Fig.~\ref{fig:maxflux}.

In the presence of background lines, extra care is needed to take them into account.  In our analysis, we conservatively take the potential dark matter line and background lines to be degenerate.  The line flux limit near the energy of a background line is therefore significantly weakened, set roughly by the actual observed line flux.  This explains the large fluctuations seen in our flux limit, especially at $E>20$\,keV, where background lines are ubiquitous.

\section{\label{sec:jfactor} Dark Matter Signal Modeling}

For a generic decaying dark matter, the expected flux from a pointed observation is
\begin{eqnarray}
\frac{dF}{dE} &=&  \frac{\Gamma} {4\pi m_\chi } \frac{dN}{dE}   {\Delta\Omega} \, {\cal{J}}\, .
\end{eqnarray}
Here, $E$ is the photon energy, $\Gamma$ is the dark matter decay rate, and $m_{\chi}$ is the dark matter mass. $dN/dE = \delta(E = m_{\chi}/2 )$ is the X-ray spectrum from dark matter decay, and $\Delta\Omega = \int_{\rm FOV}\,d\Omega\,{\cal E}$ is the average solid angle taking into account the energy-independent detector efficiencies ${\cal E}$~(see Sec.~\ref{sec:zerobounce}). ${\cal{J}}$ is the J-factor, which takes into account the dark matter distribution in the FOV. 

For sterile neutrino dark matter, its decay rate into a photon and an active neutrino~($\chi \rightarrow \gamma \nu$) depends on the mass and the mixing angle between the sterile and active neutrinos, $\sin^{2}2\theta$~\cite{Shrock:1974nd, Pal:1981rm}, as
\begin{eqnarray}
\Gamma &=&  1.38\times 10^{-32}\,{\rm s}^{-1} \smallket{\frac{\sin^{2}2\theta}{10^{-10}}} \smallket{ \frac{ m_{\chi} }{ \rm keV } }^{5} \, .
\end{eqnarray}
By using the delta function approximation for the decay photon spectrum, $dN/dE$, we have ignored the dark matter line width, which is appropriate for the energy resolution of \nustar\ (see Ref.~\cite{Speckhard:2015eva, Powell:2016zbo} for the exception).

The J-factor is the line-of-sight integral of the dark matter density, averaged over the detector FOV with detector efficiency taken into account.  
For each observation, 
\begin{eqnarray}
{\cal{J}} &=& \frac{1}{\Delta\Omega}\int_{\rm FOV}d\Omega\, {\cal E} \int_{los}d\ell\, {\rho \left[ r\left( \psi, \ell \right) \right]  }  \, ,
\end{eqnarray}
where $\rho(r)$ is the dark matter density profile, $r(\psi, \ell) = \smallket{R_{\odot}^2 + \ell^2 - 2  R_{\odot}\ell \cos{\psi} }^{1/2}$ is the galactocentric radius, $\psi$ is the opening angle from the GC, $\ell$ is the line of sight distance from the observer, and $R_{\odot} = 8\,{\rm kpc}$ is the  distance to the GC.  

For dark matter density profiles, a popular choice is the generalized Navarro-Frenk-White profile~\cite{Navarro:1996gj}, $\rho(r) \propto (r/r_{s})^{-\gamma}(1+r/r_{s})^{3-\gamma}$, where $r_{s} = 20$\,kpc is the scale radius and $\gamma$ is the density slope.  
We normalize the profile to have a local density of $\rho(R_{\odot})=0.4\,{\rm GeV\,cm^{-3}}$, as suggested by recent analyses~\cite{Piffl:2014mfa, Bienayme:2014kva, Iocco:2015xga, Pato:2015dua, 2015ApJ...814...13M, 2016MNRAS.458.3839X, 2016ApJ...817...13S}. 

The inner slope is less certain and its uncertainties must be considered because our observations are around the GC.
We therefore study several cases. 
Dark matter-only simulations favor a cuspy profile with $\gamma = 1$~\cite{Navarro:1996gj}.  We denote this case simply as the NFW profile. 
The situation becomes more complicated when baryons are added.  Ref.~\cite{Calore:2015oya}, which considered a collection of simulated galaxies~(with baryons) that best satisfy the Milky Way kinematic data, showed that the density slope is steeper~($\gamma > 1$) between about $1.5-6$\,kpc and shallower ($\gamma <1$) below 1.5\,kpc, compared to NFW~(see also Ref.~\cite{Schaller:2015mua}).  
We consider the most conservative approximation of this case by taking $\gamma = 1$ down to 1.5\,kpc, and then impose a constant density core $\rho(r<1.5\,{\rm kpc}) = \rho(1.5\,{\rm kpc})$.   
We denote this case as the coreNFW profile.  
We also check the case where the inner slope is shallower all the way to the center, $\gamma = 0.7$, and denote this as sNFW.
Finally, we consider the shallow Einasto profile~(sEIN), $\rho(r) \propto { \exp\left[-2( (r/r_{s})^{\alpha} -1 )/\alpha \right]}$, where $\alpha = 0.3$ is the shape parameter and $r_{s}$ is also 20\,kpc, in contrast to the usual~(steeper) Einasto profile with $\alpha = 0.17$~\cite{Diemand:2008in}.
Both sNFW and sEIN correspond to the conservative cases found in a collection of Galactic potential models~\cite{Pato:2015dua}. 
The sNFW case is also consistent with the lower bound found in Ref.~\cite{Hooper:2016ggc}, which constrained the inner density profile using recent dark matter mass determination of the bulge-bar region~($r \sim 1-2$\,kpc).

To combine the 12 observations, 6 for each detector, the total J-factor, $\cal J^{\rm tot}$, and the total solid angle, $\Delta\Omega^{\rm tot}$, are obtained by averaging over the exposures,
\begin{eqnarray}
{\cal J^{\rm tot}} = \frac{\sum A\,T \Delta\Omega\,{\cal J}}{\sum A\,T \Delta\Omega}\,\,\, & {\rm and} \,\,\,& { \Delta\Omega^{\rm tot}} = \frac{\sum A\,T \Delta\Omega }{\sum A\,T } \, ,
\end{eqnarray}
where $A$ is the detector area, $T$ is the effective exposure time, and the sum runs through the 12 observations shown in Table~\ref{tab:obs}.  ${ \Delta\Omega^{\rm tot}}$ is approximately $3.8\,{\rm deg^{2}}$, and ${\cal J^{\rm tot}}$ is 46, 29, 29 and, 33${\rm \,GeV\,cm^{-3}\,kpc\,sr^{-1}}$ for NFW, coreNFW, sNFW, and sEIN, respectively.  

The relatively small deviation in ${\cal J^{\rm tot}}$ due to profile uncertainties is another advantage of using the 0-bounce photons in this analysis.  The larger FOV and the blockage of the GC by the optical bench make the J-factor less sensitive to the choice of the density profiles. 
For reference, the combined J-factor corresponds to the intensity from dark matter decays at about 2$^\circ$ angle from the GC in the coreNFW profile. 
We conservatively use the coreNFW profile for our default results.

Combining all terms, the integrated photon number flux from sterile neutrino dark matter decay is 
\begin{eqnarray}\label{eqn:total_flux}
F &=&  \frac{\Gamma} {4\pi m_\chi }  \, \Delta\Omega\,{\cal{J}}\, \\
&\simeq& 2.6\times 10^{-6}\,{\rm cm^{-2}\,s^{-1} } \smallket{ \frac{m_{\chi}}{\rm 20\,keV}}^{4} \smallket{ \frac{\sin^{2}2\theta}{10^{-14}}} \nonumber \\
&&\times \smallket{ \frac{\Delta\Omega}{\rm 4\,deg^{2}}} \smallket{ \frac{\cal J}{\rm 40\,GeV\,cm^{-3}\,kpc\,sr^{-1}}} \, . \nonumber
\end{eqnarray}

There is also a contribution from extragalactic dark matter decays, but it is negligible in this case.   
For reference, integrated number flux is
\begin{eqnarray}\label{eqn:eg_flux}
{F^{\rm EG}} &=&  \frac{\Omega_{\chi} \rho_{c} } {4\pi m_\chi }\Gamma \frac{c}{H_{0}} \Delta\Omega \int dE \frac{E^{-1} \Theta(m_{\chi}/2 - E) }{\sqrt{\Omega_{\Lambda} + \Omega_{M} \smallket{  {m_{\chi}}/{2E} }^{3} } }  \nonumber \\
& \simeq & 3.6 \times 10^{-8}\,{\rm cm^{-2}\,s^{-1}} \times \smallket{ \frac{m_{\chi}}{\rm 20\,keV}}^{4} \smallket{ \frac{\sin^{2}2\theta}{10^{-14}}} \nonumber  \\
&&   \smallket{ \frac{\Delta\Omega}{\rm 4\,deg^{2}}} \smallket{ \frac{\int dE[...]}{0.1} },
\end{eqnarray}
where $\Omega_{\Lambda} = 0.685$, $\Omega_{M} = 0.315$, $\Omega_{\chi} = 0.265$, $H_{0} = 67.3\,{\rm km\,s^{-1}Mpc^{-1}}$, and $\rho_{c} = 4.26\times 10^{-6}\,{\rm GeV\,cm^{-3}}$ are the parameters in the $\Lambda$CDM cosmology~\cite{Agashe:2014kda}.  The line shape of the extragalactic component is broadened by cosmological redshifts, and is given by the integrand above.  This integral is approximately 0.1 after integrating over 10\% of the line energy at rest.

\section{\label{sec:results} Dark Matter Constraints}

\subsection{New constraint on generic dark matter}

\begin{figure}
\includegraphics[width=1.0\linewidth]{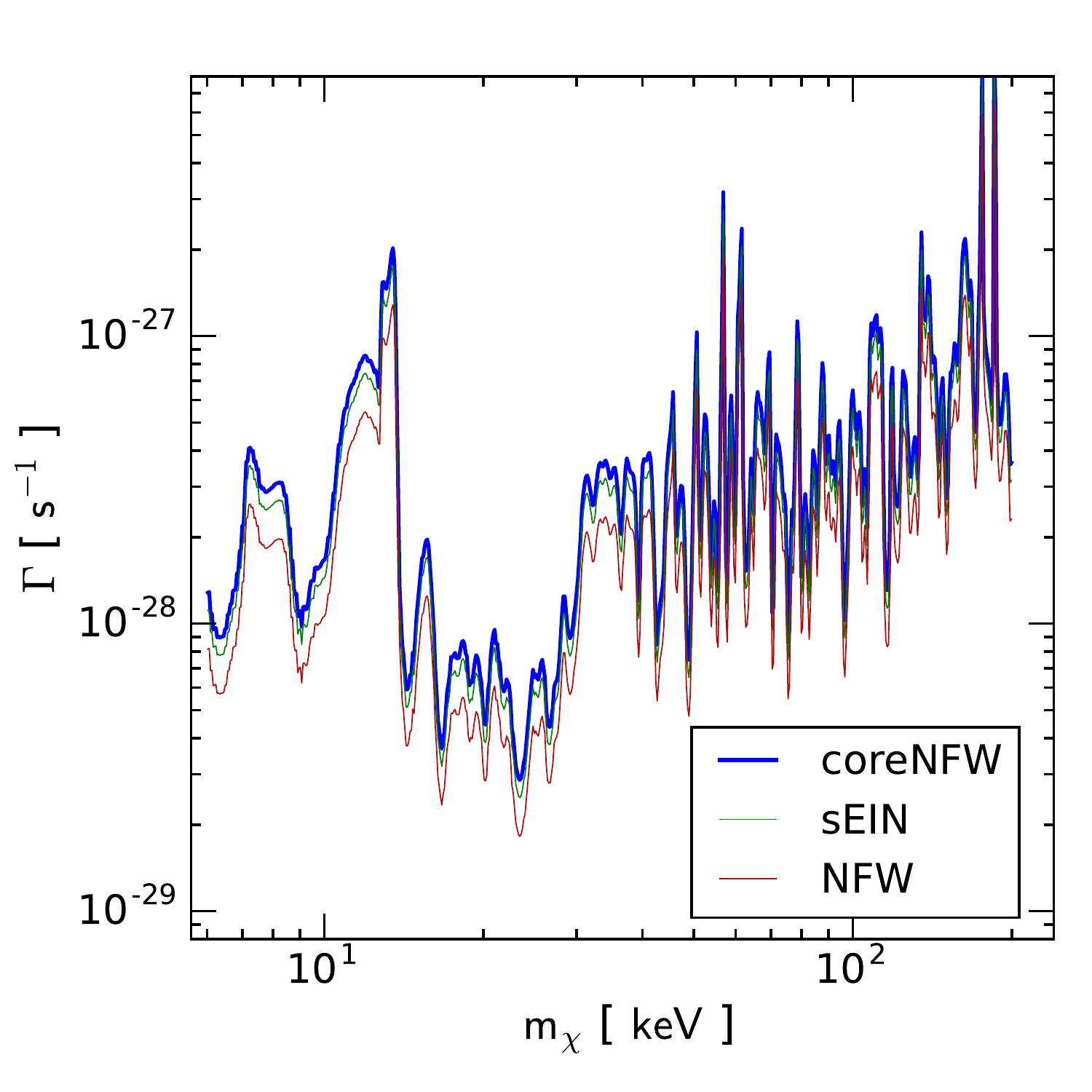}
\caption{\label{fig:decay_rate} 
Model-independent constraint on the decay rate of dark matter, assuming the final state has one monoenergetic photon with $E_\gamma = m_{\chi}/2$.  
The three curves bracket the uncertainty associated with the choice of dark matter density profile, which are all normalized to a local density value of $0.4\,{\rm GeV\,cm^{-3}}$.  Our default result uses the coreNFW profile. }
\end{figure}

In Fig.~\ref{fig:decay_rate}, we show the model-independent upper limit on the dark matter decay rate, derived using the flux upper limit shown in Fig.~\ref{fig:maxflux}, assuming a decay with one monoenergetic final state photon with $E_\gamma = m_{\chi}/2$.  This limit can be readily translated to any decaying dark matter model with a line spectrum, such as those studied in Ref.~\cite{Kusenko:2012ch, Essig:2013goa, Albert:2014hwa}.

\subsection{Summary of prior constraints on sterile neutrino dark matter}

\begin{figure}[t]
\includegraphics[width=1.0\linewidth]{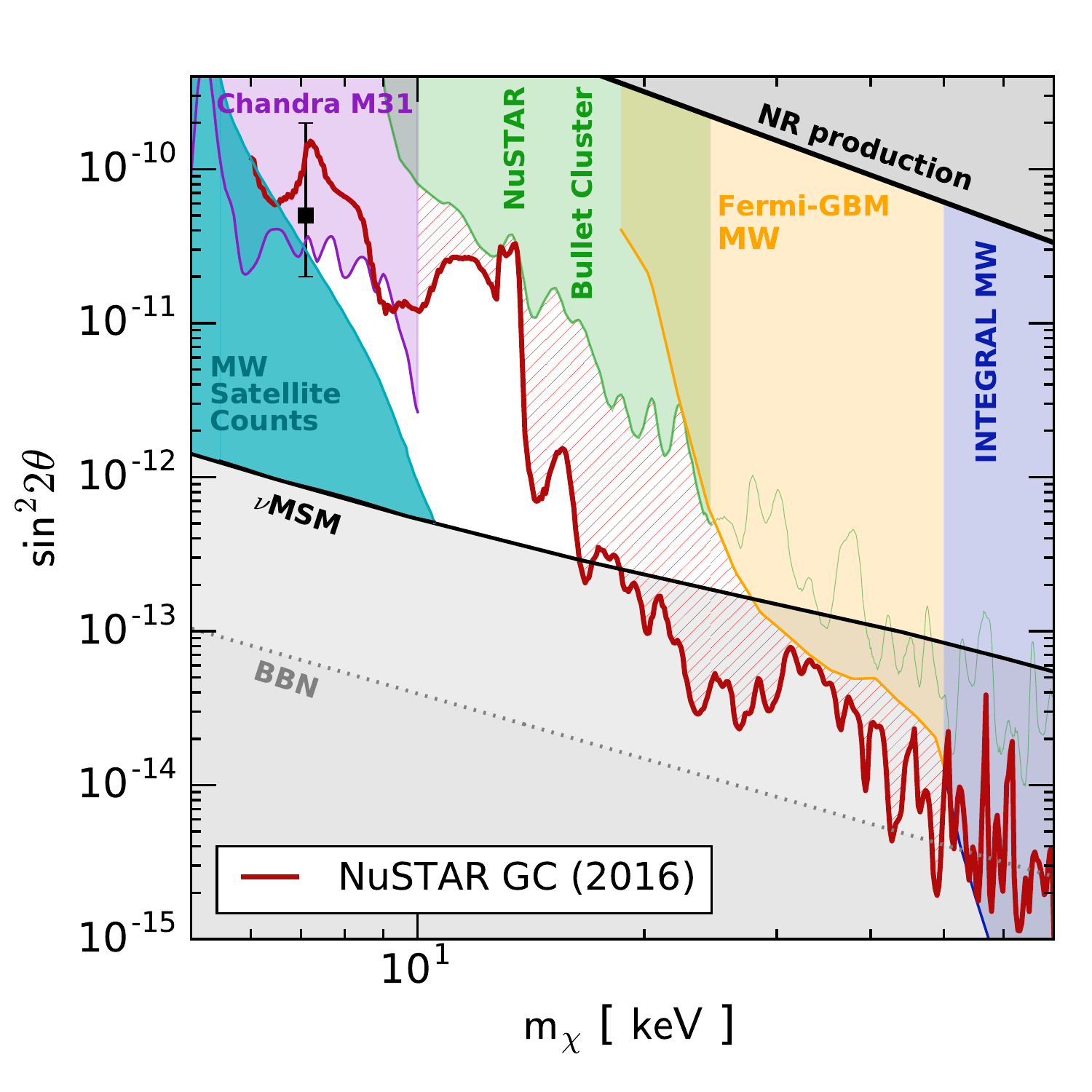}
\caption{\label{fig:constraint} 
A more detailed summary of  constraints on sterile-neutrino dark matter in the $\nu$MSM, including the constraint derived in this work. Note the changes in axis ranges from Fig.~\ref{fig:intro}.  The observed dark matter abundance can be obtained for the parameter space between the solid black lines. The upper black line corresponds to non-resonant production~(no lepton asymmetry)~\cite{Dodelson:1993je, Asaka:2006nq}.  The lower black line corresponds to resonant production with maximum lepton asymmetry in $\nu$MSM~\cite{Shi:1998km, Canetti:2012kh}; the dotted line indicates the model-independent lower bound on lepton asymmetry from BBN~\cite{Boyarsky:2009ix}.
Most of the parameter space between production constraints is ruled out by limits from structure formation~(assuming resonant production)~\cite{cherry:2016xxx} or astrophysical X-ray observations~\cite{Horiuchi:2013noa, Riemer-Sorensen:2015kqa, Ng:2015gfa, Boyarsky:2007ge}, which are now indicated individually by the colored, labelled regions.  The parameters of the tentative signal at $E\simeq $\,3.5\,keV~($m_{\chi}\simeq 7$\,keV)~\cite{Bulbul:2014sua, Boyarsky:2014jta} are shown by the black square.
Our new constraint, indicated by the red line and hatched region, rules out approximately half of the previously allowed parameter space (white region). }
\end{figure}

Here we describe the current constraints on sterile neutrino dark matter. Constraints from production mechanism and structure formation are applicable to models where sterile neutrino dark matter is produced via either resonant or non-resonant production, such as $\nu$MSM.  Importantly, astrophysical X-ray constraints are independent of the production method.
This discussion lends perspective on the importance of our result.  These limits are shown in Fig.~\ref{fig:constraint}. 

{\bf Constraints from dark matter production:} 
Sterile neutrino dark matter can be produced through a tiny mixing with active neutrinos.  In the absence of any primordial lepton asymmetry, this is known as non-resonant~(NR) production, first proposed by Dodelson and Widrow~\cite{Dodelson:1993je}.  This scenario defines an upper bound in the mass-mixing plane, above which too much dark matter would be produced.  However, in the presence of a large lepton asymmetry, the effective mixing angle would be modified by the extra matter potential, and sufficient dark matter can be produced even with a smaller mixing angle, a scenario known as resonant production, first proposed by Shi and Fuller~\cite{Shi:1998km}~(see Ref.~\cite{Venumadhav:2015pla} for the latest calculation).  In a specific model, such as $\nu$MSM~\cite{Asaka:2005an, Boyarsky:2006jm, Asaka:2006nq, Canetti:2012vf, Canetti:2012kh}, the parameter space is therefore also bounded from below, set by requiring a sufficient amount of lepton asymmetry to be generated from the model to produce the observed dark matter abundance.  If one is agnostic to the origin of the lepton asymmetry, but still requires sterile neutrino dark matter to be produced via non-resonant and resonant production, then a more model-independent and relaxed constraint can be set with Big Bang Nucleosynthesis~(BBN)~\cite{Boyarsky:2009ix}, which directly constrains the maximum amount of lepton asymmetry allowed in the Early Universe~\cite{Canetti:2012kh}.  Moreover, it is important to note that the lower bound on the mixing angle can be relaxed if sterile neutrinos were produced by other mechanisms~(for examples, see Ref.~\cite{Shaposhnikov:2006xi, Kusenko:2006rh, Merle:2013wta, Frigerio:2014ifa, Lello:2014yha, Merle:2015oja, Patwardhan:2015kga}). 

{\bf Structure formation constraints:}  
For some part of the mass range, sterile neutrinos produced via mixing can be warm dark matter.  They can suppress small-scale structure formation, which can potentially solve some of the small-scale problems seen in CDM simulations~(\cite{Weinberg:2013aya, Bozek:2015bdo, Horiuchi:2015qri}, and see Ref.~\cite{Adhikari:2016bei} and reference therein).  Conversely, structure formation can also be used to constrain the ``warmness'' of sterile neutrino dark matter, which translates roughly to the mass of sterile neutrinos.  The most robust constraint on the sterile neutrino mass can be obtain using phase space arguments, requiring $m_{\chi} \gtrsim1.7$\,keV~\cite{Horiuchi:2013noa}.  A stronger constraint can be obtained by using satellite galaxy counts~\cite{Horiuchi:2013noa,Lovell:2015psz, Schneider:2016uqi, cherry:2016xxx}. 
 Even stronger constraints may be obtained using Ly-$\alpha$ observations~\cite{Boyarsky:2008xj, Viel:2013apy, Garzilli:2015iwa, Baur:2015jsy, Schneider:2016uqi}.  However, extra care must be taken when using the Ly-$\alpha$ constraints, due to the non-negligible effect of the gas dynamics of the inter-galactic medium on the Ly-$\alpha$ signal, which can erase warm dark matter features~\cite{Viel:2013apy, Kulkarni:2015fga}. 
In this work, we conservatively adopt the galaxy counting constraint obtained by Ref.~\cite{cherry:2016xxx}, whose limit is comparable to that in Ref.~\cite{Schneider:2016uqi}. Both analyses take into account the mixing angle dependence of the mass constraint due to the different power spectrum cutoff from resonant-production calculations~\cite{Venumadhav:2015pla}.  

{\bf X-ray limits:}  
The radiative decay of sterile neutrinos allows astrophysical observations to set upper limits on the mixing angle~\cite{Dolgov:2000ew, Abazajian:2001vt}.  These limits are independent of the production mechanism of sterile neutrinos.
For $m_{\chi}\lesssim 10$\,keV, strong constraints have been obtained using \chandra, \suzaku, and \xmm~\cite{ Boyarsky:2006zi,RiemerSorensen:2006fh,Abazajian:2006yn,Watson:2006qb, Boyarsky:2006ag, RiemerSorensen:2006pi, Abazajian:2006jc, Boyarsky:2006kc, Boyarsky:2007ay, Loewenstein:2008yi, RiemerSorensen:2009jp, Loewenstein:2009cm, Mirabal:2010an, Watson:2011dw, Borriello:2011un, Horiuchi:2013noa, Riemer-Sorensen:2014yda, Sekiya:2015jsa}.   
For $m_{\chi}\gtrsim$\,50\,keV, strong constraints have been obtained using \integral~\cite{Yuksel:2007xh, Boyarsky:2007ge}, completely ruling out mixing-produced sterile neutrinos as the sole dark matter constituent. 
For 10\,keV $\lesssim m_{\chi}\lesssim 50$\,keV, limits have been set by \emph{HEAO-1}~\cite{Boyarsky:2005us, Boyarsky:2006fg}, \emph{Fermi-GBM}~\cite{Ng:2015gfa}, and \nustar~\cite{Riemer-Sorensen:2015kqa}~(observations of the Bullet Cluster using focused photons).  We summarize the overall X-ray limits, using results from Refs.~\cite{Horiuchi:2013noa, Riemer-Sorensen:2015kqa, Ng:2015gfa, Boyarsky:2007ge}, in Fig.~\ref{fig:constraint}. 

\subsection{New constraint on sterile neutrino dark matter}

In Fig.~\ref{fig:constraint}, we show the limit obtained with our analysis, together with the existing constraints mentioned above.  Near $m_{\chi} = 20$\,keV, our result improves the limit by about one order of magnitude and significantly reduces the remaining parameter space.  
This does not imply that now $\nu$MSM is less likely to be a viable theory of nature, because only a single point in the parameter space is sufficient to realize the theory.  However, it does mean that the model is closer to being completely tested.

Compared to previous limits set using \nustar\ observations of the Bullet Cluster~\cite{Riemer-Sorensen:2015kqa}, our results are stronger, mainly due to the close proximity of the GC and the large dark matter mass enclosed by our 0-bounce solid-angle aperture.
Assuming the $\nu$MSM framework, our limit translates into an upper limit on the sterile neutrino mass of $m_{\chi} \lesssim 16$\,keV.  

For $m_{\chi}$ near 13\,keV and 40\,keV, the deterioration in the limits~(``bumps'') are associated with photon energies where there is strong astrophysical iron line emission and where the GXRE spectrum transitions into the detector background spectrum, respectively.  

For $m_{\chi}<$\,10\,keV, our limit becomes worse than that of Ref.~\cite{Horiuchi:2013noa}, due to the presence of lines near 3.5\,keV and 4.5\,keV whose nature is not totally clear~\cite{Wik:2014boa}.  Interestingly, a tentative signal at 3.5\,keV was discovered with other instruments~\cite{Bulbul:2014sua, Boyarsky:2014jta}, which can potentially be explained by a sterile neutrino at $m_{\chi} \simeq 7$\,keV~\cite{Abazajian:2014gza, Venumadhav:2015pla}; however, its origin is still heavily debated~\cite{Jeltema:2014qfa, Boyarsky:2014ska, Anderson:2014tza, Urban:2014yda, Gu:2015gqm, Jeltema:2015mee, Ruchayskiy:2015onc, Bulbul:2016yop, Aharonian:2016gzq}, and may require the next generation instruments~\cite{Iakubovskyi:2014yxa, Neronov:2015kca} or novel dark matter detection techniques~\cite{Speckhard:2015eva, Powell:2016zbo} to settle the case.

To help elucidate the nature of the 3.5\,keV line in our data set, we consider a small part of the observations where the FOV is blocked by the Earth.  Both 3.5\,keV and 4.5\,keV lines are found in the occulted data set with consistent strengths as the GC data set.  They are not as significant as in the GC observations, but the statistics in the occulted data is also lower.  This reinforces the interpretation of these lines being detector backgrounds of \nustar .
The determination of their nature, however, is beyond the scope of this work.

\subsection{Towards closing the $\nu$MSM sterile neutrino window}

For sterile neutrino dark matter in $\nu$MSM, only a tiny window remains near $m_{\chi} \simeq 10-16$\,keV.  Unfortunately, our analysis at this energy is hampered by the strong astrophysical iron line. 
In the future, the sensitivity could be improved by using observations of fields with weaker astrophysical emission, or by improving the astrophysical and detector background modeling. 

In addition, improved sensitivity to warm dark matter can be achieved in the future with new surveys of satellite galaxies~\cite{cherry:2016xxx}, [58], or with new methods of probing dark matter subhalos~\cite{Hezaveh:2016ltk, Bovy:2016irg}. 
Together with new X-ray observations, new warm dark matter studies, and new limits on sterile neutrinos from supernovae~\cite{Arguelles:2016uwb}, the full parameter space of sterile neutrino dark matter in the $\nu$MSM can soon be fully explored.  In the case of a null detection, it will further motivate sterile neutrino dark matter models with other production mechanisms~\cite{Shaposhnikov:2006xi, Kusenko:2006rh, Merle:2013wta, Frigerio:2014ifa, Lello:2014yha, Merle:2015oja, Patwardhan:2015kga}.  This also means that physics in addition to the minimal assumption in $\nu$MSM is needed to explain dark matter, baryon asymmetry, and neutrino mass.

\section{\label{sec:conclusion} Conclusions}

We search for dark matter that decays into monoenergetic keV-scale photon lines using a subset of the \nustar\ Galactic plane survey data.  No obvious dark matter signals are found, and thanks to the novel use of 0-bounce photons, robust and stringent upper limits are placed on the decay rate of dark matter into X-rays.  Our analysis has produced the strongest indirect detection limit on dark matter lines in the energy range $E_\gamma = 5-25$\,keV. 

This also allows us to place strong upper limits on the mixing angle for sterile neutrino dark matter.  For the $\nu$MSM, where the sterile neutrino is produced via mixing in the Early Universe, only a small section of the original parameter space remained before out work. Our results significantly reduce the available parameter space, which is likely to be completely probed by future analyses of \nustar\ observations.  
In the case of a null detection, it would imply the minimalistic approach of $\nu$MSM is insufficient to explain neutrino mass, baryon asymmetry, and dark matter simultaneously.
It would also further heighten interest in models where sterile neutrino dark matter is produced with different mechanisms~\cite{Shaposhnikov:2006xi, Kusenko:2006rh, Merle:2013wta, Frigerio:2014ifa, Lello:2014yha, Merle:2015oja, Patwardhan:2015kga}.

\bigskip
{\bf Note added:} As this paper was being completed, we learned of a sterile neutrino dark matter search that also considered the use of \nustar\  0-bounce photons, but with a different data set~\cite{Neronov:2016wdd}. Our limit is comparable, and in some cases, more stringent, than theirs. 

\section{\label{sec:acknowledgement} Acknowledgements} 
We thank Kevork Abazajian, Alexey Boyarsky, Jaesub Hong, Marco Drewes, Dima Iakubovskyi, Denys Malyshev, Alexander Merle, Signe Riemer-Sorensen, Oleg Ruchayskiy, and Daniel Wik for helpful discussions.
We thank Signe Riemer-S\o rensen for data files from  Ref.~\cite{Riemer-Sorensen:2015kqa}. 
KP and CH thank Joe Cammisa for his generous help with NuSTAR software infrastructure. 
KCYN was supported by the Ohio State University Presidential Fellowship and NSF Grant PHY-1404311.  JFB is supported by NSF Grant PHY-1404311. CH was supported by the Haverford College KINSC Summer Scholar program.  RK acknowledges support from Russian Science Foundation~(grant 14-22-00271).

\bibliographystyle{h-physrev}
\bibliography{references}

\end{document}